# SCALING OF PSEUDO-CRITICAL COUPLINGS

# IN TWO-FLAVOUR QCD

Frithjof Karsch

HLRZ, c/o KFA Jülich, P.O. Box 1913, D-52425 Jülich, Germany;
Fakultät für Physik, Universität Bielefeld, P.O.Box 100131, D-33501 Bielefeld, Germany;*
Institute for Theoretical Physics, University of California, Santa Barbara, CA 93106-4030.

## Abstract

We study the scaling behaviour of the pseudo-critical couplings for the chiral phase transition in two-flavour QCD. We show that all existing results from lattice simulations on lattices with temporal extent $N_\tau = 4$, 6 and 8 can be mapped onto a universal scaling curve. The relevant combination of critical exponents, $\beta\delta$, is consistent with the scaling behaviour expected for a second order phase transition with $O(4)$ exponents. At present, scaling according to the $O(2)$ symmetry group can, however, not be ruled out.

*Permanent address

The chiral phase transition in two flavour QCD has been studied extensively over the last years on lattices of varying spatial and temporal extent [1-8]. So far no indications for a possible first order transition have been found suggesting that this transition is, in the limit of vanishing quark masses, a second order phase transition. If this transition is already controlled by the restoration of the continuum $SU(2) \times SU(2)$ chiral symmetry, one would expect that the critical behaviour can be described in terms of universal properties in the vicinity of the critical point of three dimensional, $O(4)$ symmetric spin models [9-11]. In lattice studies of QCD with staggered fermions the situation gets complicated due to the fact that the lattice action has only a $U(1) \times U(1)$ chiral symmetry, which in the strong coupling limit also leads to a second order finite temperature phase transition with $O(2)$ critical exponents [9].

The critical exponents for $O(2)$ and $O(4)$ spin models are quite similar. It thus will remain difficult for quite some time to distinguish these symmetry groups in numerical studies of the chiral phase transition in two-flavour QCD with quarks of mass $m_q$, only through a determination of the critical exponents. Additional information can, however, be obtained by examining in how far physical observables scale according to the expected continuum version of the relevant fields, ie. reduced temperature $t$ and external (magnetic) field $h$,

$$t = \frac{T - T_c}{T_c} \quad , \quad h = \frac{m_q}{T} \quad . \tag{1}$$

This implies, that the temperature and quark mass dependence of physical observables studied in the vicinity of the critical point, $(T = T_c, m_q = 0)$, on lattices with different values of the cut-off should allow for a common description in terms of the above continuum parameters. We will discuss here to what extent such a scaling behaviour is fulfilled by the presently available data. In particular, we will examine the scaling behaviour of the pseudo-critical couplings, determined from simulations at non-zero values of the quark masses. We will show, that these pseudo-critical couplings scale in terms of the lattice version of the scaling variables $t$ and $h$. The combination of critical exponents $(\beta\delta)$, controlling the quark mass dependent shifts of these couplings, is at present consistent with the expected behaviour for $O(4)$ symmetry restoration as well as with scaling according to $O(2)$ exponents. We will discuss the calculation of further observables, which should lead to more accurate location of the pseudo-critical couplings and thus a better determination of critical exponents.

In the case of the temperature driven second order phase transition in SU(2) gauge theory it could be shown that the critical behaviour of the continuum theory can be deduced from a unified finite size scaling analysis of simulations at different values of the lattice cut-off [12]. Such a finite size scaling analysis cannot directly be taken over to the case of QCD, where the correlation length at the critical point is limited due to the explicit symmetry breaking introduced through the finite quark mass rather than the finite size of the system [9]. Finite size effects are thus suppressed for non-vanishing values of the quark masses. This opens the possibility to study directly the influence of the symmetry breaking external fields without having to worry about strong contaminations of the scaling behaviour due to finite lattice

size effects.[1]

In the presence of an external symmetry breaking field, $h$, the singular part of the free energy has the scaling property

$$f(t,h) = b^{-1} f(b^{y_t} t, b^{y_h} h), \qquad (2)$$

where $b$ is an arbitrary scale factor, $y_t$ and $y_h$ are the thermal and magnetic critical exponents. In the case of an $O(4)$ symmetric theory in three dimensions one has for instance, $\delta = y_h/(1-y_h) = 4.82(5)$ and $\beta = (1-y_h)/y_t = 0.38(1)$ [13], while the corresponding exponents for an $O(2)$ symmetric theory are $\delta = 4.755(6)$ and $\beta = 0.3510(2)$. In the case of QCD the reduced temperature $t$ and the external field $h$ are given in terms of the gauge coupling $g^2$ and the bare quark mass, $ma$, on a lattice of temporal extent $N_t$ as

$$\begin{aligned} t &= \frac{6}{g^2} - \frac{6}{g_c^2(0)} \;, \\ h &= maN_\tau \;. \end{aligned} \qquad (3)$$

Here $g_c^2(0)$ denotes the critical coupling in the limit of vanishing bare quark mass. For non-vanishing values of the quark mass a pseudo-critical coupling $g_c^2(ma)$, may, for instance, be defined as the location of a peak in the magnetic susceptibility, $\chi = \frac{\partial^2}{\partial h^2} f(t,h)$. From eq. (2) one finds for the location of this peak,

$$t_{max} \sim h^{1/\beta\delta} \;. \qquad (4)$$

The pseudo-critical couplings in two-flavour lattice QCD have so far been studied on lattices of size $N_\tau \times N_\sigma^3$ with $N_\tau = 4$, 6 and 8 and various values of $ma$ [1-8]. Typically the spatial size of the lattice has been chosen to be twice as large as the temporal extent, $N_\sigma \simeq 2N_\tau$. Studies of finite size effects suggest that at least for the purpose of locating the pseudo-critical couplings this seems to be sufficiently large for the presently used range of quark masses [3, 6, 7].

So far the pseudo-critical couplings have been extracted by determining the point of largest slope of the order parameter $\langle \bar{\psi}\psi \rangle$ as a function of $6/g^2$. A direct calculation of the derivative $\chi = \frac{\partial}{\partial h} \langle \bar{\psi}\psi \rangle$ at non-vanishing $6/g^2$ has not yet been performed, although in the strong coupling limit $(6/g^2 \equiv 0)$ this has proven to be very helpful in locating the pseudo-critical couplings [9].[2]

In the absence of such calculations, the point of largest curvature can only be estimated from the slope of $\langle \bar{\psi}\psi \rangle$. In fig. 1a we show the relevant data available for lattices of temporal extent $N_\tau = 8$. Our estimates for the pseudo-critical couplings, together with those published for $N_\tau = 4$ and 6 [1-8] are given in the Table. We note that our estimate for the pseudo-critical coupling for $N_\tau = 8$, $ma = 0.0125$ differs somewhat from the previously published value [7]. For this value of the quark mass we show data obtained from calculations on lattices with spatial extent $N_s = 12$ [6] and 16 [7]. Note that no lattice size dependence is visible in the data. We also have added a data point at $\beta = 5.40$ to this curve taken from a simulation on a $6 \times 12^3$ lattice [5]. This seems to be legitimate for our purpose of determining the largest slope in $\langle\bar\psi\psi\rangle$ as a calculation for $N_\tau = 8$ at this value could only lead to a slightly larger value. A smooth interpolation of this combined data sample leads to the estimate for the pseudo-critical coupling given in the Table. For $ma = 0.004$ some indications for a two-state signal had been reported [8], which, however, disappeared in a large volume ($8 \times 32^3$) simulation at the relevant value of the gauge coupling, $6/g^2 = 5.48$ [15]. This suggests that fluctuations on the $8 \times 16^3$ lattice are large at this small value of the quark mass. In fig. 1a we have used an average of the data given in Ref. 8 for $\langle\bar\psi\psi\rangle$ at $\beta = 5.48$, $ma = 0.004$ [15].

---

[1] The lattice size independence of the pseudo-critical couplings for the currently used quark masses on lattices with temporal extend $N_\tau = 4$ has been verified in Ref. 3. For $N_\tau = 8$, $ma = 0.0125$ the same conclusion can be drawn from the data published in Refs. 6 and 7.

[2] The derivative of $\langle \bar{\psi}\psi \rangle$ has been calculated in the quenched approximation at non-vanishing values of $6/g^2$ [14]. There it has been used to improve the extrapolation of the chiral condensate to vanishing quark masses.

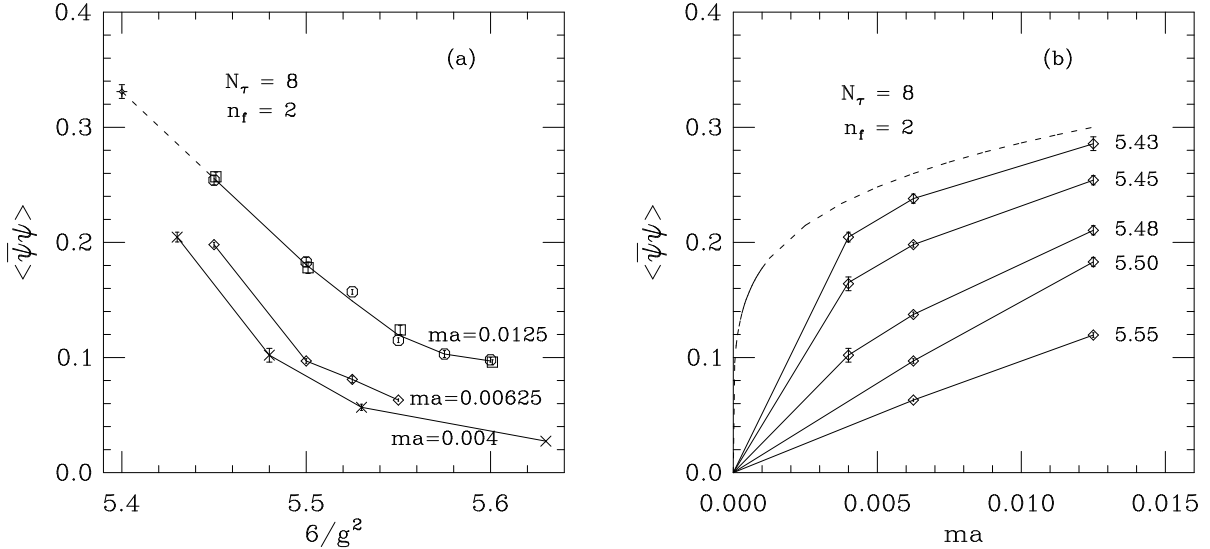

Figure 1: The chiral order parameter on lattices with temporal extent $N_\tau = 8$. Details on the data are given in the text. Solid lines are drawn to guide the eye. The dashed curve in fig. 1b is the expected behaviour for a magnetic phase transition with $O(4)$ critical indices, $\langle\bar\psi\psi\rangle(6/g_c^2(0), ma) \sim ma^{1/\delta}$. A curve with the $O(2)$ exponents would only differ insignificantly from this.

If the pseudo-critical couplings reflect the expected scaling behaviour of continuum QCD, their quark mass dependence should be described by

$$\frac{6}{g_c^2(ma, N_\tau)} = \frac{6}{g_c^2(0, N_\tau)} + c(maN_\tau)^{1/\beta\delta}, \qquad (5)$$

with a universal constant $c$, which in leading order is independent of $N_\tau$, $ma$ as well as $t$.

|  | $6/g_c^2(ma, N_\tau)$ | | |
| --- | --- | --- | --- |
| $ma$ | $N_\tau = 4$ | $N_\tau = 6$ | $N_\tau = 8$ |
| 0.004 | – | – | 5.450 (30) |
| 0.00625 | – | – | 5.475 (25) |
| 0.01 | 5.265 (5) | – | – |
| 0.0125 | 5.270 (2) | 5.420 (10) | 5.490 (30) |
| 0.025 | 5.288 (2) Ref. 3 | 5.445 (5) | – |
| 0.025 | 5.291 (1) Ref. 4 | – | – |
| 0.05 | 5.320 (10) | 5.470 (40) | – |
| 0.1 | 5.375 (10) | 5.525 (40) | – |

Table: Pseudo-critical couplings on various size lattices and for various values of the bare quark mass [1-8]. We note that the numbers for $N_\tau = 8$ are based on a re-evaluation of existing data for the chiral order parameter $\langle\bar\psi\psi\rangle$ as discussed in the text.

The pseudo-critical couplings are shown in fig. 2 as a function of $h = maN_\tau \equiv m_q/T$. The data for $N_\tau = 4$ have been fitted to the scaling form, eq. (5), assuming critical exponents for a three dimensional, $O(4)$ symmetric model, i.e. $1/\beta\delta = 0.546$.

This two-parameter fit fixes the universal slope parameter $c$, which then is used in single parameter fits to the $N_\tau = 6$ and 8 data sets.

This determines the critical couplings in the $ma \to 0$ limit. We find

$$c = 0.231(20) \tag{6}$$

and

$$\frac{6}{g_c^2} = \begin{cases} 5.225(5) , & N_\tau = 4 \\ 5.363(4) , & N_\tau = 6 \\ 5.424(16) , & N_\tau = 8 \end{cases} \tag{7}$$

The fact that all pseudo-critical couplings for $N_\tau = 4$, 6 and 8 can be fitted to the form given by eq. (5) with a unique coefficient $c$ implies that they all follow a single scaling curve when plotted in terms of the reduced external field variables $t$ and $h$ defined in eq. (3). This is shown in fig. 3. We stress, however, that this analysis can, at present, not distinguish the $O(4)$ exponents from a possible scaling according to the $O(2)$ symmetric form of the staggered lattice action. In this case one would have $1/\beta\delta = 0.599$, which would allow an equally good fit. We also have performed fits allowing for the exponent $1/\beta\delta$ as a free parameter. Such a five-parameter fit to the complete sample of pseudo-critical couplings from simulations for $N_\tau = 4$, 6 and 8 yields $1/\beta\delta = 0.57 \pm 0.16$. We note, however, that scaling of the pseudo-critical couplings for different values of $N_\tau$ according to the scaling fields defined in eq.(3) only is expected to hold in a regime of couplings close to the continuum limit, it certainly does not hold in the strong coupling limit.

A further check on the consistency of the analysis in terms of a second order phase transition comes through the investigation of the quantity

$$\Delta(t,h) = \frac{h\chi}{\langle\bar\psi\psi\rangle} . \tag{8}$$

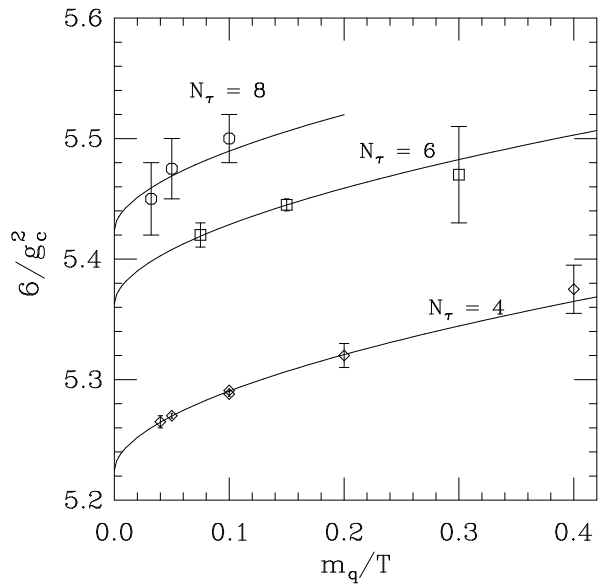

Figure 2: Pseudo-critical couplings versus $m_q/T$ on lattices of temporal extent $N_\tau = 4$, 6 and 8. The curves are fits to the data using eq. (5) with $O(4)$ critical exponents as explained in the text. Using $O(2)$ exponents would allow for fits of similar quality.

It has been pointed out in Ref. 9 that this quantity has properties very similar to those of order parameter cumulants examined usually in finite size scaling studies of temperature driven second order phase transitions. In particular curves for different, but fixed, values of $h$ all cross at $t = 0$ in a unique point, $\Delta(0, h) = 1/\delta$. At least in principle this allows a unique determination of the zero-field critical coupling as well as the critical exponent $\delta$. Here we want to show that, given the value of $\delta$, eq.(7) yields values for the zero-field critical couplings which are consistent with our previous analysis. We have calculated some approximants for $\Delta$, using differences of the order parameter calculated at neighbouring values of $ma$ to determine the derivative $\chi$ (fig. 1b). Using data on the $N_\tau = 8$ lattices for $ma = 0.004$ and $0.00625$, we find

$$\Delta(t, \overline{ma} = 0.005125) = \begin{cases} 0.34(8), & 6/g^2 = 5.43 \\ 0.43(4), & 6/g^2 = 5.45 \\ 0.67(3), & 6/g^2 = 5.48 \end{cases} . \qquad (9)$$

Although the errors are, at present, large for this quantity, a linear fit suggests that $\Delta$ will take on values close to 0.2 for $\beta \simeq 5.42$, which is consistent with the result from our fits given in eqs.(6) and (7). A direct evaluation of the derivative $\chi$ should allow to reach a much higher accuracy.

We have analyzed all existing data on the pseudo-critical couplings for the chiral phase transition in two-flavour QCD assuming a second order phase transition. We have shown that all data for the pseudo-critical couplings are naturally explained by this assumption and show scaling behaviour in terms of the variable $h \equiv maN_\tau$, expected to hold close to the continuum limit. At present our analysis certainly does not give stringent bounds on the scaling exponent $1/\beta\delta$. In particular we cannot rule out scaling according to the $O(2)$ symmetry of the staggered fermion action. However, we want to stress that the present analysis suggests that there is no significant change in the pseudo-critical behaviour of two-flavour QCD when going from lattices with temporal extent $N_\tau = 4$ to $N_\tau = 8$. More detailed studies on $N_\tau = 4$ lattices along the line presented here, should allow for a more precise determination of critical exponents than it was possible here.

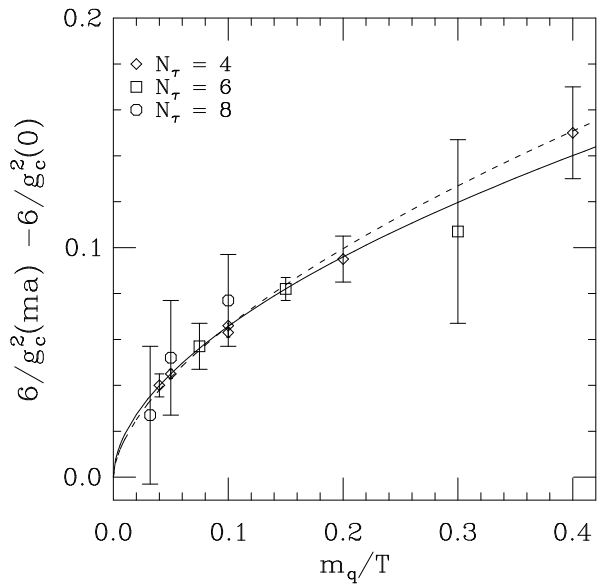

Figure 3: Scaling plot of the pseudo-critical couplings. The solid curve is a fit with the $O(4)$ exponent, $1/\beta\delta = 0.546$, while the dashed curve shows a fit with the $O(2)$ exponent $1/\beta\delta = 0.599$.

## Acknowledgements

I very much appreciated discussions with Steve Gottlieb, Urs Heller, Bob Sugar and Doug Toussaint. I would like to thank them as well as Norman Christ for making available to me their data collections on the two-flavour QCD phase transition. This research was supported in part by the National Science Foundation under Grant No. PHY89-04035 and the German Research Foundation, DFG, under Grant, Pe340/3-2.